\documentclass[reprint,amsfonts,amsmath,amssymb,aps,pra,twocolumn]{revtex4-2}
\usepackage[utf8]{inputenc}
\usepackage{graphicx}
\usepackage{physics}
\usepackage{dcolumn}
\usepackage{bm}
\usepackage{color}
\usepackage{ulem}

\begin{document}

\title{State distinguishability under weak measurement and post-selection: A unified system and device perspective}
\author{Philipp Stammer}
\email{stammer@mbi-berlin.de}
\affiliation{Max-Born-Institute for Nonlinear Optics and Short Pulse Spectroscopy, Berlin, Germany}
\affiliation{Technische Universität Berlin, Institute for Theoretical Physics, Berlin, Germany}

\date{\today}

\begin{abstract}

We quantify the disturbance of a quantum state undergoing a sequence of observations, and particularly focus on a weak measurement followed by post-selection and compare these results to the projective counterpart.
Taking into account the distinguishability of both, the system and the device, we obtain the exact trade-off between the system state disturbance and the change of the device pointer state. 
We show that for particular post-selection procedures the coupling strength between the system and the device can be significantly reduced without loosing measurement sensitivity, which is directly transferred to a reduced state disturbance of the system. 
We observe that a weak measurement alone does not provide this advantage but only in combination with post-selection a significant improvement in terms of increased measurement sensitivity and reduced state disturbance is found. 
We further show that under realistic experimental conditions this state disturbance is small, whereas the exact post-selection probability is considerably larger than the approximate value given by the overlap of the initial and final state when neglecting the system state disturbance.

\end{abstract}

\maketitle

\section{\label{sec:intro}Introduction}

Suppose we perform a sequence of observations of the observables $A$, and then $B$, on an initially prepared system described by the state $\rho_i$ \cite{feynman48}. In the final $B$ measurement we select only those cases where the measurement corresponding to a particular outcome is affirmative, say this outcome is described by the eigenstate $\ket{f}$. 
In such an experiment the observable $A$ is measured on the pre- and post-selected (PPS) sub-ensemble defined by the initial preparation and final measurement.

To consistently describe this measurement as an inherent active process the system under investigation has to interact with a measuring device \cite{peres2006quantum}, and correlations between both systems will emerge. 
For a strong coupling we have the projective von Neumann measurement \cite{neumann2013mathematische, luders1950zustandsanderung} and the conditional expectation value of the measured observable $A$ on the PPS ensemble follows from the Aharonov-Bergmann-Lebowitz (ABL) formula \cite{aharonov1964time}

\begin{equation}
\expval{A}_{ABL} = \sum_\nu a_\nu \frac{{\Tr(\Pi_f \Pi_{A_\nu} \rho_i \Pi_{A_\nu})} } { {\sum_\mu \Tr(\Pi_f \Pi_{A_\mu} \rho_i \Pi_{A_\mu})}},
\end{equation}

where $\Pi_f = \dyad{f}$ is the projector on the final state, and $a_\nu$ and $\Pi_{A_\nu} = \dyad{\chi_\nu}$ are the eigenvalues and the projector on the eigenstates of $A$, respectively .

On the contrary, if the coupling is sufficiently weak, the outcome of such a weak measurement \cite{wiseman2009quantum} on a PPS ensemble is given by the weak value (WV) \cite{aharonov1988result, ritchie1991realization}

\begin{align}
\label{eq:intro_def_wv}
\expval{A}_{WV} = \sum_\nu a_\nu \frac{\Tr(\Pi_f \Pi_{A_\nu} \rho_i)}{\sum_\mu \Tr(\Pi_f \Pi_{A_\mu}\rho_i)} = \frac{\bra{f}A \rho_i \ket{f}}{\bra{f}\rho_i\ket{f}}.
\end{align}

The measurement of the WV was subject of an extensive discussion \cite{duck1989sense, wiseman2002weak,  dressel2014colloquium, vaidman2017weakcontroversy}.The ability to measure the WV was leading to an analysis of fundamental quantum aspects \cite{steinberg1995much, williams2008weak, mir2007double, lund2010measuring, kocsis2011observing, rozema2012violation, pan2020interference} and to explain the observed phenomena \cite{lundeen2009experimental, aharonov2013classical, danan2013asking, chatzidimitriou2019weak}. The application of WV to quantum state measurement \cite{lundeen2011direct, lundeen2012procedure, thekkadath2016direct, wu2013state} is still of current interest \cite{xu2020direct} and WV are extensively used for quantum metrology \cite{dixon2009ultrasensitive, hofmann2011uncertainty, PhysRevLett.125.080501}. 

However, to measure the WV and to extract information from the system is by no means possible without disturbing it \cite{fuchs1998information}. 
The trade-off between the measurement disturbance and the information gain has been investigated to a large extend in the context of quantum cryptography \cite{fuchs1996quantum, fuchs2001information}. In the context of weak and protective measurements it has been shown that such measurements cause a minimal disturbance of the system \cite{aharonov1990properties, schlosshauer2015analysis}. 
Thus, in WV measurement protocols the disturbance of the system during the intermediate weak measurement is usually neglected \cite{aharonov1990properties} and quantified by the transition probability from the initial to an orthogonal state \cite{schlosshauer2015analysis, ipsen2015disturbance}.
From the perspective of the gained information the analysis is usually focused on the (quantum) Fisher information of the measurement outcome \cite{tanaka2013information, knee2014amplification, zhang2015precision, harris2017weak, PhysRevLett.125.080501}, and no direct relation to the corresponding state disturbance is provided.

In the following we will present a unified analysis of the state disturbance of the system and the change of the device state, accounting for the information gain, for a generic weak value experiment. 
This is done by quantifying the distinguishability between the states during the measurement sequence. This also takes into account the change to non-orthogonal states, important for quantum state differentiation \cite{ivanovic1987differentiate, peres1988differentiate}. 
With the exact model used to describe the WV measurement protocol, we can quantify the mutual state change of the system and the device during the measurement sequence. This allows to directly relate the information gain of the device in the trade-off to the system state disturbance.
By quantifying the system state disturbance in a WV measurement protocol will help to analyze the boundaries of the experimental setup and to examine the approximations made when predicting experimental outcomes. For instance, we will show that the system state disturbance for a specific measurement protocol is small, whereas the exact post-selection probability can be significantly larger than the approximate value which neglects this state disturbance.

This paper is organized as follows. In Sec. \ref{sec:measurement} we introduce the exact model to describe the measurement sequence. 
With the exact expressions we investigate the disturbance of the system state during the measurement sequence in Sec. \ref{sec:distance_example}. We compare a weak measurement with the projective counterpart and include a post-selection procedure.
In Sec. \ref{sec:Fidelity_device} we analyze the change of the device state during the measurement sequence and discuss the relation to the weak value amplification technique. 
In Sec. \ref{sec:minimal_disturbance} we explicitly connect the mutual change of the system and device state to obtain the exact trade-off relation between the system state disturbance and the change of the device pointer state. We further discuss the assumption of negligible state disturbance during a weak measurement and show that the resulting approximate expression for the post-selection probability can considerably differ from the exact value under realistic experimental conditions. In Sec. \ref{sec:conclusion} we conclude the paper.

\section{\label{sec:measurement}Measurement sequence and conditional expectation value}

In this section we will introduce the model to describe the generalized measurement, where the internal degree of freedom of a system is coupled to a variable in configuration space. This coupling is quite general \cite{bray2020dissecting} and we will later specify the generic case of a two-level system coupled to its coordinate, e.g. a Stern-Gerlach type experiment.
The following analysis provides an exact model for the measurement sequence which extends the description in terms of state vectors from \cite{zhu2011quantum} to density matrices. We can thus describe the exact mixed state of the system originating from the measurement.

Consider an ensemble of systems described by an initial state $\rho_i$ and a measurement device which is prepared in a well defined state $\sigma_i = \dyad{\phi}$. The measurement interaction Hamiltonian couples the system observable $A$ to the device pointer momentum $p$ with canonical conjugate pointer position $q$, such that  $[q,p] = i$ ($\hbar =1$) 

\begin{align}
\label{hamiltonian}
H_I (t) = g \delta(t-t_0) A \otimes p,
\end{align}

where $g$ is the coupling strength and $\delta(t-t_0)$ account for an impulsive measurement interaction at time $t_0$ \cite{jozsa2007complex, dressel2015weak}. Prior to the measurement the system and the device are in an uncorrelated product state $\Upsilon_i = \rho_i \otimes \sigma_i$ and during the measurement evolve to the entangled state $\Upsilon^\prime = U (\rho_i \otimes \sigma_i ) U^\dagger$, with the unitary evolution given by $U = \exp(-i g A \otimes p)$. The total state after the measurement interaction is 

\begin{align}
    \Upsilon^\prime = \sum_{\nu \mu} \bra{\chi_\nu} \rho_i \ket{\chi_\mu} \dyad{\chi_\nu}{\chi_\mu} e^{-i g a_\nu p} \dyad{\phi} e^{i g a_\mu p},
\end{align}

where $a_\nu$ and $\ket{\chi_\nu}$ are the eigenvalues and eigenstates of $A$, respectively.

To obtain the pointer reading of the device $\expval{q} = \Tr(q \Upsilon^\prime)$, we specify the initial device state to be a Gaussian centered at $q=0$ with spread $\Delta$ in the $q$-representation $\phi(q) \equiv \braket{q}{\phi} = \left(2 \pi \Delta^2\right)^{-1/4} \exp[- q^2/(4 \Delta^2)]$. Thus the pointer position after the measurement interaction, $\expval{q} = g \expval{A}_i$, is proportional to the expectation value of the system observable in the initial state $\expval{A}_i = \Tr_S(A \rho_i)$, where $\Tr_S(\cdot)$ denotes the trace over the system degrees of freedom.

Now, suppose the final post-selection is performed on the system by a projective measurement of some other observable, and we select on those outcomes corresponding to the eigenstate $\ket{f}$, i.e. $\Upsilon_f = \Pi_f \Upsilon^\prime \Pi_f / \Tr(\Pi_f \Upsilon^\prime)$.

The total state after such a post-selection is 

\begin{align}
\label{eq:total_state_postselection}
    \Upsilon_f =& \frac{1}{P(f)} \sum_{\nu \mu} \bra{\chi_\nu} \rho_i \ket{\chi_\mu} \braket{f}{\chi_\nu} \braket{\chi_\mu}{f}  \\
    & \times  \dyad{f} e^{-i g a_\nu p} \dyad{\phi} e^{ig a_\mu p}, \nonumber
\end{align}

where $P(f) = \Tr(\dyad{f} \Upsilon^\prime)$ is the post-selection probability for measuring the property corresponding to the state $\ket{f}$. 
Accordingly, the corresponding pointer shift is given by 

\begin{align}
\label{eq:expval_exact}
    \expval{q}_{i, f} =& \frac{g}{P(f)} \sum_{\nu \mu} \braket{f}{\chi_\nu} \braket{\chi_\mu}{f} \mel{\chi_\nu}{\rho_i}{\chi_\mu}  \\
    & \times \frac{1}{2} \left( a_\nu + a_\mu \right) e^{- \frac{g^2}{8 \Delta^2}(a_\nu - a_\mu)^2}, \nonumber 
\end{align}

which is the exact conditional expectation value of $A$ for a generalized measurement \cite{duck1989sense, zhu2011quantum}. The post-selection probability is 

\begin{equation}
\label{eq:post-selection-probability}
P(f) = \sum_{\nu \mu} \braket{f}{\chi_\nu} \braket{\chi_\mu}{f} \mel{\chi_\nu}{\rho_i}{\chi_\mu} e^{-g^2(a_\nu - a_\mu )^2/(8 \Delta^2)}.
\end{equation}

In the strong coupling limit, $g \gg \Delta$, the conditional pointer reading reduces to the ABL formula $\expval{q}_{ABL} = g {\sum_\nu a_\nu \abs{\braket{f}{\chi_\nu}}^2 \bra{\chi_\nu}\rho_i \ket{\chi_\nu} }/{\sum_\mu \abs{\braket{f}{\chi_\mu}}^2 \bra{\chi_\mu} \rho_i \ket{\chi_\mu}} $. 

However, in the weak measurement limit, $g \ll \Delta$, the exponential in \eqref{eq:total_state_postselection} can be expanded to first order \cite{aharonov1988result, aharonov1990properties}, and we obtain the familiar result for a weak measurement of $A$ on a PPS ensemble

\begin{align}
\label{eq:expval_WV}
    \expval{q}_{WV} = g \operatorname{Re} \left( \frac{ \mel{f}{A \rho_i}{f} }{\mel{f}{\rho_i}{f}} \right),
\end{align}

given by the real part of the WV \eqref{eq:intro_def_wv}.
Note that the exact outcome \eqref{eq:expval_exact} reproduces the WV \eqref{eq:expval_WV} for $g/ \Delta \to 0$, i.e. $\lim_{g/\Delta \to 0} \expval{q}_{i, f} = \frac{1}{2} \frac{g}{P(f)} \left[ \mel{f}{A \rho_i}{f} + \mel{f}{\rho_i A}{f} \right] = \expval{q}_{WV}$, where we have used that $\lim_{g/\Delta \to 0} P(f) = \mel{f}{\rho_i}{f}$. While the shift in the pointer position is proportional to the real part of the WV, the imaginary part is imprinted in a change of the pointer momentum \cite{jozsa2007complex}. 

To specify the model we will compare the different measurement protocols for a Stern-Gerlach type experiment, where the system observable is given by the $z$-component of the spin of a spin-1/2 particle, i.e. $A = \sigma_z$, and is coupled to the particle's momentum \eqref{hamiltonian}. 
Initially we prepare the system in the spin-up state along the $\xi$-direction with angle $\alpha$ to the $x$-axis in the $x$-$z$-plane, $\rho_i = \dyad{\uparrow_\xi}$ with $\ket{\uparrow_\xi} = \cos\alpha/2 \ket{\uparrow_x} + \sin \alpha/2 \ket{\downarrow_x}$. To define the PPS ensemble we select for final spin-up states along the $\eta$-direction with angle $\beta$ to the $x$-axis, i.e. $\ket{\uparrow_\eta} = \cos \beta /2 \ket{\uparrow_x} + \sin \beta /2 \ket{\downarrow_x}$. The initial preparation and the final post-selection are likewise performed with a Stern-Gerlach device along the $\xi$- and $\eta$-direction, respectively.

For this particular example the exact result of the conditional expectation value is given by 

\begin{align}
\label{eq:expval_special_exact}
    \expval{\sigma_z}_{i, f} = \frac{\sin \alpha + \sin \beta }{1 + \sin \alpha \, \sin \beta + e^{- \frac{g^2}{2 \Delta^2}} \cos \alpha \, \cos \beta }.  
\end{align}

This expression reproduces the conditional expectation value following ABL $ \expval{\sigma_z}_{ABL}= {(\sin \alpha + \sin \beta)}/{(1 + \sin \alpha \, \sin \beta)}$ and the weak value $\expval{\sigma_z}_{WV} = {\sin \left[ \frac{1}{2} (\alpha + \beta)\right]}/{\cos \left[ \frac{1}{2} (\alpha - \beta)\right]}$, for $g/ \Delta \gg 1$ and $g / \Delta \ll 1$, respectively. 

\begin{figure}
    \centering
    \includegraphics[width=7.5cm]{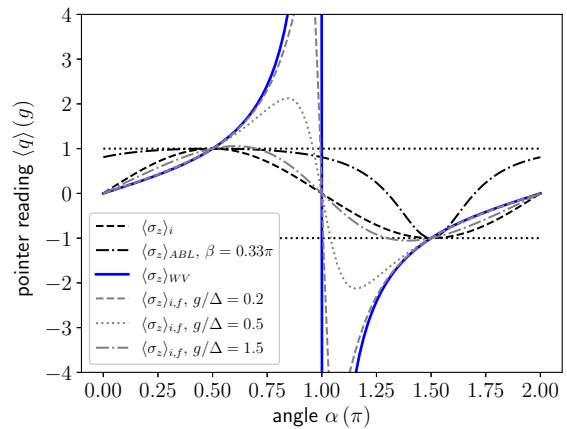}
      \caption{Pointer reading $\expval{q}$ for different measurement protocols of the $z$-component of the spin of a spin-1/2 particle for different preparation procedures with angle $\alpha$. The final post-selected state is $\ket{\uparrow_x}$ with $\beta = 0$. The expectation value in the initial state $\expval{\sigma_z}_i = \sin \alpha$ (black, dashed) is bounded by $\pm 1$ (dotted horizontal lines). The conditional expectation value following the ABL formula is shown for $\beta = 0.3 \pi$ (black, dash-dot) and obeys the same boundaries like the expectation value. The peculiarity of the WV (blue, solid) arise when it exceeds the conventional boundaries of projective measurements. The exact result for different order parameters $g / \Delta$ recover the ABL and the WV in the corresponding limits (gray).}
      \label{fig:expectation_values}
\end{figure}

The measurement outcomes for different orientations of the initial state are shown in Fig. \ref{fig:expectation_values}, where we have taken $\beta = 0$ \cite{note2}.
The conventional expectation value (without post-selection)  $\expval{\sigma_z}_i = \sin \alpha$, is bounded by $-1 \le \expval{\sigma_z}_i \le 1$. 
Similarly the conditional expectation value following ABL lies within the same bounds.

The striking feature in the weak measurement regime, when $\expval{\sigma_z}_{i,f} \ge \expval{\sigma_z}_i$, is due to the non-vanishing overlap of the device states which cause quantum interference \cite{dressel2015weak}. 
As $\alpha \to \pi $ the weak value diverges since the initial ($\ket{\downarrow_x}$) and final state ($\ket{\uparrow_x}$) become orthogonal. 

\section{\label{sec:distance_example}Distinguishability during measurement sequence}

We shall now investigate the effect of the measurement sequence on the system state and quantify the disturbance caused by a generalized measurement including post-selection. 
As a measure of this disturbance we will use the distinguishability between two states. 
The advantage of this measure is that we can use it to define a distance between the two states. This allows to quantify all changes of the system state and we are not limited to quantify the disturbance in terms of the transition to an orthogonal state. In doing so, we are also able to identify the influence of the measurement on the coherence of the system.

\subsection{Comparing strong and weak measurement}

The initial preparation procedure generates spin-up states $\ket{\uparrow_\xi} = \cos \alpha/2 \ket{\uparrow_x} + \sin \alpha /2 \ket{\downarrow_x}$ (all density matrices are represented in the measured $\sigma_z$ basis)

\begin{align}
\rho_i =     \begin{pmatrix}
 \frac{1}{2} (1+ \sin \alpha) & \frac{1}{2} \cos \alpha\\
 \frac{1}{2} \cos \alpha & \frac{1}{2} (1- \sin \alpha)
\end{pmatrix}.
\end{align}

For the intermediate projective measurement (without post-selection), which will destroy all coherence, the state is left in a statistical mixture of eigenstates $\ket{\uparrow_z}$ and $\ket{\downarrow_z}$ 

\begin{align}
    \rho_s = \begin{pmatrix}
 \frac{1}{2} (1+ \sin \alpha) & 0\\
 0 & \frac{1}{2} (1- \sin \alpha)
\end{pmatrix}.
\end{align}

On the contrary the weak measurement only disturbs the initial state slightly, such that the exact system state is given by

\begin{align}
\label{eq:matrix_intermediate_weak}
    \rho_w &= \sum_{\nu \mu \in \{ \uparrow_z, \downarrow_z  \}} \bra{\chi_\nu} \rho_i \ket{\chi_\mu} \dyad{\chi_\nu}{\chi_\mu} e^{- \frac{g^2}{8 \Delta^2} (a_\nu - a_\mu)^2} \nonumber \\
    & = \begin{pmatrix}
 \frac{1}{2} (1+ \sin \alpha) & \frac{1}{2} \cos \alpha \, e^{- \frac{g^2}{2 \Delta^2} }\\
 \frac{1}{2} \cos \alpha \, e^{- \frac{g^2}{2 \Delta^2}} & \frac{1}{2} (1- \sin \alpha)
\end{pmatrix}.
\end{align}

Note that following the original derivation of the WV in \cite{aharonov1988result, aharonov1990properties}, the decoherence factor $\exp[-g^2 /(2 \Delta^2)]$ vanish and is identical with \eqref{eq:matrix_intermediate_weak} when $g/\Delta \to 0$.

To quantify the disturbance of the system during the measurement sequence we use the trace distance between to density matrices $D(\rho_1,\rho_2) = \frac{1}{2} \Tr(\abs{\rho_1 - \rho_2})$ as a measure of how much the system state after the measurement deviates from the state prior to the interaction. The trace distance has the properties of a metric in the space of density matrices and a natural interpretation as the distinguishability between the two states \cite{fuchs1999cryptographic, nielsen2002quantum}. 
For a spin-1/2 particle the trace distance between two states $\rho_1$ and $\rho_2$ with matrix elements $(\rho_{1/2})_{ij}$ has the simple form $D(\rho_1, \rho_2) = \sqrt{a^2 + \abs{b}^2}$, where $a = (\rho_1)_{11} - (\rho_2)_{11}$ and $b = (\rho_1)_{10} - (\rho_1)_{10}$, are the differences in the population and coherence of the two density matrices, respectively.

At first, we consider the trace distance between the initial state $\rho_i$ and the intermediate state after a strong or weak measurement. For a strong von Neumann measurement we have $D(\rho_i,\rho_s) = \frac{1}{2} \abs{\cos \alpha}$, while for the weak measurement

\begin{align}
\label{eq:distance_initial_stron_vs_weak}
D(\rho_i,\rho_w) &= \frac{1}{2} \abs{\cos \alpha} \left(1- e^{-g^2/(2 \Delta^2)}\right)\\
& \le D(\rho_i,\rho_s), \nonumber
\end{align}

demonstrating that the distinguishability between the state prior to the measurement and the state after the measurement is smaller for the weak measurement case than in the projective counterpart.
Interpreting $D(\rho_i, \rho_{s/w})$ as the amount the system state changes, or equivalently, the amount of disturbance caused by the measurement, 
confirm that a weak measurement is minimal disturbing. For $g/\Delta \to 0$, such that $D(\rho_i,\rho_w) = 0$, the initial state is undisturbed, but no information is obtained. In the opposite limit $g/\Delta \to \infty$ we recover the effect of a projective measurement, leading to an enhanced information gain in the trade-off for a larger state disturbance \cite{fuchs2001information}. 
Both measurements are non-disturbing for $\alpha \in \{ \pi / 2 ,\, 3 \pi/2\}$, which is the case for a procedure which initially prepares an eigenstate of the measured observable. 

\begin{figure}
    \centering
    \includegraphics[width=6.5cm]{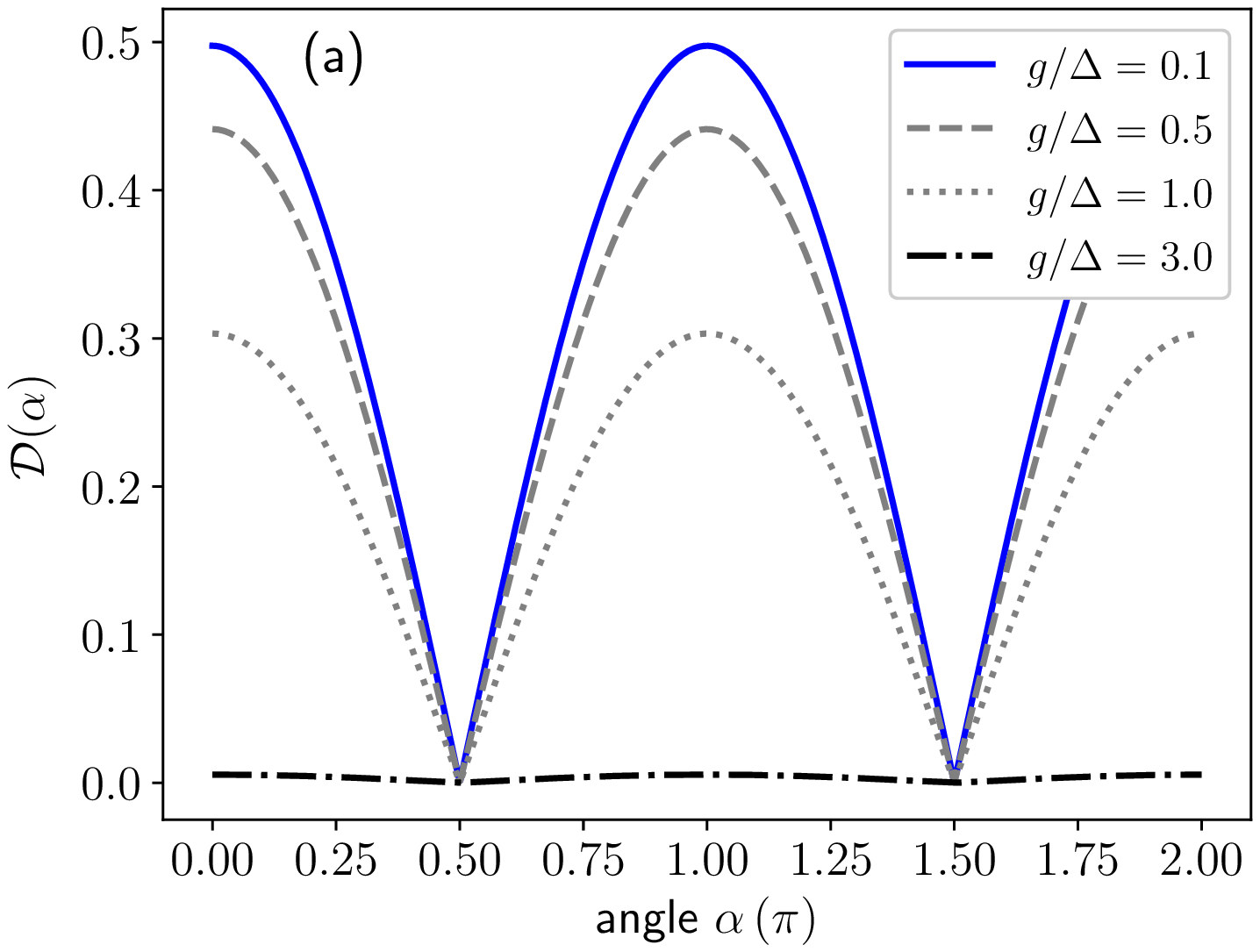}
    \includegraphics[width=6.5cm]{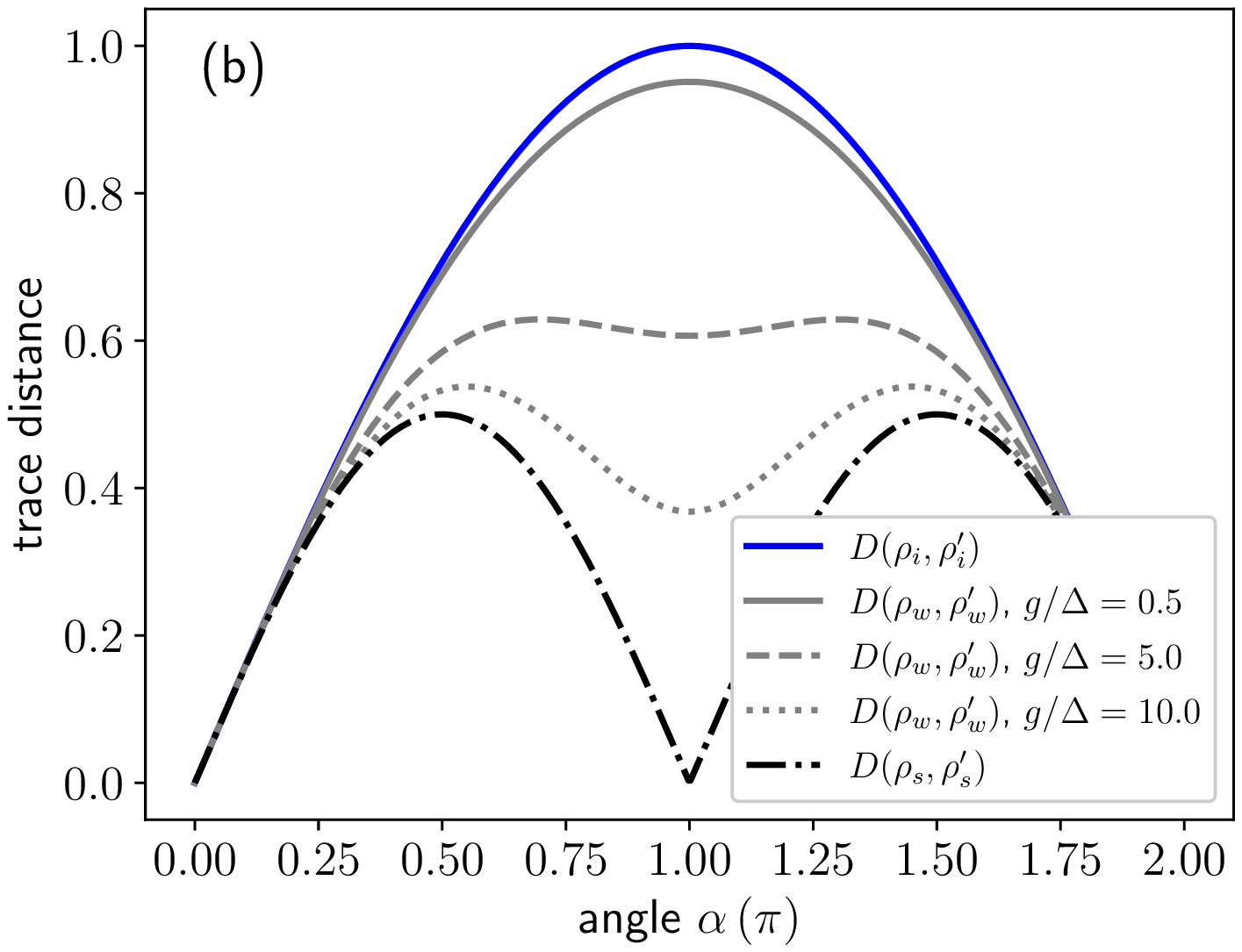}
      \caption{(a) Measure of preserved information in the system (attributed to preserved coherence) of a weak measurement compared to a projective measurement $\mathcal{D}(\alpha) = \frac{1}{2} \abs{\cos \alpha} \exp[- g^2 / (2 \Delta^2)]$ for different initial preparation angles $\alpha$ and order parameters $g / \Delta$. (b) Distinguishability of two different initial preparations $\rho_i$ and $\rho_i^\prime$ prior to the measurement $D(\rho_i, \rho_i^\prime)$ (blue, solid) and following a weak measurement on both ensembles $D(\rho_w, \rho_w ^\prime)$ (gray) or projective measurement $D(\rho_s, \rho_s^\prime)$ (black, dash-dot). In all cases we have fixed $\alpha^\prime =0$.}
      \label{fig:distance_weak_vs_strong}
\end{figure}

To quantify the difference between both experimental schemes in terms of the corresponding amount of preserved coherence (the populations remain unaltered), we show the difference of the trace distance of both schemes $\mathcal{D}(\alpha) = D(\rho_i,\rho_s) - D(\rho_i,\rho_w) $ for different values of the order parameter $g/\Delta$ in Fig. \ref{fig:distance_weak_vs_strong} (a). This is a direct measure of the preserved coherence in a weak measurement process and is associated with the conservation of information inside the system during a weak measurement. 
For larger $\mathcal{D}(\alpha)$, and thus smaller $D(\rho_i,\rho_w)$, the amount of preserved coherence increase and is maximal for $\alpha \in \{ 0, \, \pi \}$ when the initial state $ \rho_i =  (\ket{\uparrow_z} \pm \ket{\downarrow_z})(\bra{\uparrow_z} \pm \bra{\downarrow_z}) / 2$ is an equal superposition of the eigenstates corresponding to the measured observable $\sigma_z$. 
We observe that $\mathcal{D}(\alpha) = \frac{1}{2} \abs{\cos \alpha} \exp[- g^2 / (2 \Delta^2)] \ge 0$, implying that the strong measurement remove more information carried by the system than a weak measurement, but in return increase the information gain.
This illustrate the fact that a weak measurement is less disturbing and will be used to obtain the trade-off relation between the information gain and the system disturbance \cite{fuchs1996quantum, fuchs2001information} in the following sections.

\subsection{Distinguishability of two ensembles}

The minimal disturbing characteristic of weak measurement does not only leads to a survival of coherence in face of performing a measurement, but additionally help to preserve the distinguishability of two different ensembles. 
Suppose we prepare two different initial states $\rho_i$ and $\rho_i^\prime$ with angles $\alpha$ and $\alpha^\prime$, respectively. Any measurement ultimately leads to a reduced distinguishability of these two ensembles due to the contraction property of the trace distance \cite{nielsen2002quantum}. However, if the measurement is weak, the distinguishability between the two ensembles will never be smaller than for a projective measurement  

\begin{align}
    &D(\rho_i,\rho_i^\prime) = \sqrt{\frac{1}{2}\left[1-\cos(\alpha - \alpha^\prime)\right]}  \\
    &\ge D(\rho_w, \rho_w^\prime)  \nonumber \\
    &= \frac{1}{2} \sqrt{(\sin \alpha - \sin \alpha^\prime)^2 + (\cos \alpha - \cos \alpha^\prime)^2e^{- g^2 / \Delta^2}}  \nonumber \\
     &\ge D(\rho_s, \rho_s^\prime) = \frac{1}{2} \abs{\sin \alpha - \sin \alpha^\prime}.  \nonumber
\end{align}

The inequality $D(\rho_w, \rho_w^\prime) \ge D(\rho_s, \rho_s^\prime)$ has a geometrical interpretation on the Bloch sphere. Recalling that after the measurement the two ensembles of two-level systems are described by mixed states $\rho = \frac{1}{2} [\vb{1} + \vb{r} \cdot \bm{\sigma}]$, where $\vb{1}$ is the identity matrix, the bold $\bm{\sigma}$ is the vector of Pauli matrices and $\abs{\vb{r}}^2 \le 1$. In fact we have $\abs{\vb{r}_s}^2 = \sin^2 \alpha$, and $\abs{\vb{r}_w}^2 = \sin^2 \alpha + \cos^2 \alpha \, \exp[- g^2/\Delta^2] $, and hence $\abs{\vb{r}_s}^2 \le \abs{\vb{r}_w}^2$. Thus the two mixed states after a strong measurement are each located on a sphere with smaller radius than for the weak measurement case. 

The dependence of the distinguishability between two ensembles for the different measurement schemes, compared to the initial distinguishability, can be seen in Fig. \ref{fig:distance_weak_vs_strong}(b) (we have fixed the preparation of the second ensemble to $\alpha^\prime = 0$). 
The behavior of $D(\rho_i,\rho_i^\prime)$ (blue, solid) is expected, since the two ensembles have a maximal distinguishability when the two states are orthogonal for $\alpha = \pi$. For these orthogonal ensembles the distinguishability after the projective measurement (black, dash-dot) has vanished since the measurement of $\sigma_z$ will produce identical mixed states. To enhance the distinguishability of the two ensembles compared to a projective measurement, we can perform a weak measurement (gray) in order to minimally disturb each ensemble and retain their distance.

Note that this distinguishability between two ensembles is more robust against stronger coupling than the disturbance of a single ensemble. In Fig. \ref{fig:distance_weak_vs_strong} (a) a weak measurement for an order parameter of $g / \Delta = 3$ is virtually equivalent to a projective measurement in terms of the state disturbance, whereas for the distinguishability of two ensembles in Fig. \ref{fig:distance_weak_vs_strong}(b) there is still a significant distinction in the distinguishability between a projective and a weak measurement for $g / \Delta = 10$. 

The property of enhanced state distinguishability between the two mixed states following a weak measurement compared to the projective counterpart is directly manifested in the enhanced probability of correct state identification $P = \frac{1}{2} [1+ D(\rho_1, \rho_2)]$, with the trace distance as the bias in favor of correct state identification for an optimal measurement strategy \cite{nielsen2002quantum}. We thus have $P_w = \frac{1}{2} [1 + D(\rho_w, \rho_w^\prime)] \ge P_s$. 
This could be used to identify the possibly unknown interaction strength $g$. Suppose two parties prepare orthogonal initial states, i.e. $D(\rho_i, \rho_i^\prime) = 1$, and send these states through the same experiment with unknown coupling strength. An observer receiving $\rho_w$ or $\rho_w^\prime$ with equal probability, can infer the coupling strength $g = \Delta \sqrt{2 \abs{\ln (2 P_w - 1)}}$, from the obtained success probability $P_w$ of state identification following the unknown interaction.

\subsection{The influence of post-selection}

We shall now investigate the consequence of the post-selection measurement on the system, which projects on the spin-up state along the $\eta$-direction, with angle $\beta$ to the $x$-axis. The system after the projective post-selection measurement is described by

\begin{align}
    \rho_f = \begin{pmatrix}
 \frac{1}{2} (1+ \sin \beta) & \frac{1}{2} \cos \beta\\
 \frac{1}{2} \cos \beta & \frac{1}{2} (1- \sin \beta)
\end{pmatrix}.
\end{align}

First, we note that due to the post-selection the two different initial preparations $\rho_i$ and $\rho_i^\prime$ will be found in the identical final state $\rho_f = \dyad{\uparrow_\eta}$, independent of the previous measurement strength. 
Thus, from the perspective of the system state the two initial ensembles can not be distinguished anymore, i.e. $ D (\rho_f, \rho_f^\prime) = 0$. And by solely observing the final state of the system no inference about the type of the previous measurement can be achieved, whereas this information is encoded in the probe state disturbance \cite{vaidman2017weak}. 

In order to investigate the effect of the post-selection on the ensemble in more detail, we compare the post-selected final state $\rho_f$ with the case if the final measurement is non-selective. The state after the non-selective final measurement is obtained when re-mixing all different outcomes $\rho_f$, weighted with the probability $p_f$ that the corresponding measurement outcome was obtained. The state if then given by $\bar \rho = \sum_f p_f \rho_f $, and will contain dependencies of the initial preparation and the type of the previous measurement imprinted in the probability of success $p_f$. 
The trace distance between the initial state and the final non-selective state $\bar \rho$ for a general intermediate measurement is

\begin{figure}
    \centering
    \includegraphics[width=6.5cm]{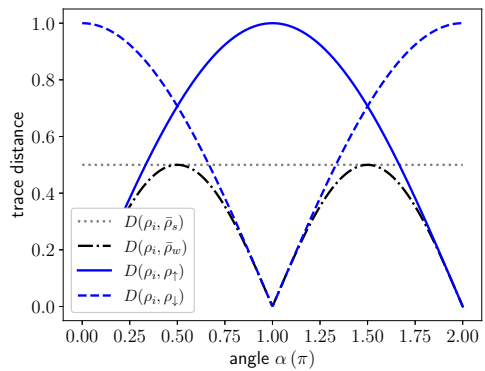}
      \caption{Distinguishability of the initial state $D(\rho_i, \bar \rho_{s,w})$ with the state after a non-selective final measurement of the spin in the $\eta$-direction for an intermediate pojective (gray, horizontal) or weak measurement (black, dash-dot). For comparison we show the distance of $\rho_i$ with the post-selected states $\rho_\uparrow = \dyad{\uparrow_\eta}$ (blue, solid) and $\rho_\downarrow = \dyad{\downarrow_\eta}$ (blue, dashed) given by $D(\rho_i, \rho_{\uparrow, \downarrow}) = \sqrt{\frac{1}{2} [1 \mp \cos(\alpha - \beta)]}$. We have set $\beta = 0$.}
      \label{fig:distances_non_selective}
\end{figure}

\begin{widetext}
\begin{equation}
D(\rho_i, \bar \rho) = \frac{1}{2} \sqrt{ \left[ A(\alpha,\beta) -  B(\alpha,\beta) \, e^{-g^2/(2 \Delta^2)} \right]^2 + \left[ \cos \alpha + B(\alpha + \pi/2, \beta) - A(\alpha+ \pi/2,\beta) \, e^{-g^2/(2\Delta^2)}  \right]^2 },
\end{equation}
\end{widetext}

where $A(\alpha, \beta) = \sin \alpha \cos^2 \beta$ and $B(\alpha,\beta) = \cos \alpha \sin \beta \cos \beta$.
In the strong and weak limit this simplifies to $D(\rho_i, \bar \rho_s) = \frac{1}{2} \sqrt{A(\alpha, \beta)^2 + [\cos \alpha + B(\alpha + \frac{\pi}{2}, \beta)]^2 }$, and $D(\rho_i, \bar \rho_w) = \frac{1}{2} \abs{\sin(\alpha-\beta)}$, respectively. We found that $D(\rho_i, \bar \rho_s) \ge D(\rho_i, \bar \rho_w)$.
This is in agreement with the inequality of Eq.\eqref{eq:distance_initial_stron_vs_weak} and the corresponding geometrical interpretation, that due to the increased disturbance of projective measurements, the states traverse a longer path on the Bloch sphere.
The distance measure in the strong $D(\rho_i, \bar \rho_s)$ (gray, horizontal) and weak $D(\rho_i,\bar \rho_w)$ (black, dash-dot) limit can be seen in Fig. \ref{fig:distances_non_selective}. For comparison we show the distance of the initial state with the two possible post-selected states $D(\rho_i, \rho_\uparrow) = \sqrt{\frac{1}{2} [1- \cos (\alpha - \beta)]}$ (blue, solid) and $D(\rho_i, \rho_\downarrow) = \sqrt{\frac{1}{2} [1+ \cos (\alpha - \beta)]}$ (blue, dashed). We observe that $D(\rho_i, \rho_{\uparrow, \downarrow}) \ge D(\rho_i, \bar \rho_w)$, which imply that the distinguishability of the initial state and any post-selected final state is always larger than the initial state with the non-selective final state following a weak measurement. This inequality holds for arbitrary initial preparation $\alpha$ and final post-selection $\beta$. This demonstrates that the effect of post-selection is decisive if it follows a weak measurement, and can considerably increase the distinguishability to the initial state in comparison to the non-selective measurement. In contrast, for a previous projective measurement such a general inequality can not be found for arbitrary initial preparations. 

To analyze the consequence on the state distinguishability caused by performing a post-selection in more general, we consider an arbitrary two-level system. The system is prepared in the pure state $\rho_i = \dyad{i}$ and post-selected on the final state $\rho_f = \dyad{f}$. The non-selective final state is $\bar \rho = p_f \dyad{f} + p_\perp \dyad{f_\perp} $, with the probability of success $p_f$ ($p_\perp$) for the post-selection $\rho_f$ (and the orthogonal state $\rho_\perp = \dyad{f_\perp}$). 
The difference in the distinguishability for the selective and non-selective final measurement obeys

\begin{align}
D(\rho_i, \rho_f) - D(\rho_i, \bar \rho) \ge  D(\rho_i, \rho_f) - \sum_f p_f D(\rho_i, \rho_f),
\end{align} 
 
where we used that $D(\rho_i, \sum_j p_j \rho_j) \le \sum_j p_j D(\rho_i, \rho_j)$. Since $p_\perp = 1- p_f$, we further have 

\begin{align}
D(\rho_i, \rho_f) &- D(\rho_i, \bar \rho) \ge \left( 1- p_f \right) \left( \abs{\braket{f_\perp}{i}} - \abs{\braket{f}{i}} \right).
\end{align}

Since $0 \le p_f \le 1$, we find that $D(\rho_i, \rho_f)  \ge D(\rho_i, \bar \rho)$ if $\abs{\braket{f_\perp}{i}}  \ge \abs{\braket{f}{i}}$. Thus, when the projection of the initial state onto the state orthogonal to the final post-selected state is larger than the projection onto the post-selection itself, than the process of post-selection is leading to an enhanced distinguishability compared to the non-selective case. This holds for arbitrary measurements and is in accordance with Fig. \ref{fig:distances_non_selective}.   
However, the inequality for the intermediate weak measurement $D(\rho_i, \rho_f)  \ge D(\rho_i, \bar \rho_w)$ holds for arbitrary initial states in a WV measurement protocol, but in general not if the previous measurement is projective.

We shall now relate the observed enhancement of the state distinguishability between the initial state and the different final states with the corresponding measurement outcome.
We have seen that any post-selected sub-ensembles in a WV measurement protocol increase the distinguishability to the initial state compared to the non-selective case. 
For the case of post-selecting $\rho_\uparrow = \dyad{\uparrow_x}$, the distance to the initial state $\rho_i = \cos \alpha / 2 \ket{\uparrow_x} + \sin \alpha / 2 \ket{\downarrow_x}$ is larger than between $\rho_i$ and the orthogonal state $\rho_\downarrow = \dyad{\downarrow_x}$, i.e. $D(\rho_i, \rho_\uparrow) \ge D(\rho_i, \rho_\downarrow)$, for $\alpha \in [\pi/2, 3 \pi/2]$. Namely, when the projection of the initial state to the state orthogonal to the post-selection is larger than the projection to the post-selection itself. 
This is the same region of initial preparations $\alpha$, where the WV $\expval{\sigma_z}_{WV} = \tan(\alpha/2)$ exceed the eigenvalue range (see Fig. \ref{fig:expectation_values}).
Note that in the region of anomalous WV the contribution from the coherence part $\abs{b}^2 = (\rho_i)_{10} - (\rho_{\uparrow})_{10}= \frac{1}{4} ( \cos \alpha + 1)^2$ in the trace distance $D(\rho_i, \rho_\uparrow)$ exceeds the contribution from the population $a^2 = (\rho_i)_{11} - (\rho_{\uparrow})_{11}= \frac{1}{4} \sin^2 \alpha$.

\section{\label{sec:Fidelity_device}Change of the pointer state}

In the previous section we have quantified the state disturbance of the system during the measurement sequence. We shall now analyze the change of the pointer state with respect to the initial state $\sigma_i =  \dyad{\phi}$, represented by $\braket{q}{\phi} \propto \exp[-q^2/(4\Delta^2)]$.
To quantify the change we use the fidelity as a distance measure between two states $F(\sigma_1, \sigma_2) = \Tr(\sqrt{\sqrt{\sigma_1} \sigma_2 \sqrt{\sigma_1}})$ \cite{fuchs1999cryptographic}. 
Since the initial pointer state is pure, the fidelity simplify to $F(\sigma_i, \sigma_2) = \bra{\phi}\sigma_2 \ket{\phi}$.

\subsection{Pointer state during measurement sequence}

If we perform a strong von Neumann measurement the fidelity with the initial state vanish $F(\sigma_i,\sigma_s) = 0$, since the post-measurement state $\sigma_s = \sum_\nu \mel{\chi_\nu}{\rho_i}{\chi_\nu} \dyad{q-g a_\nu}$, and the initial state are orthogonal (for $a_\nu \neq 0$).
The more interesting case is given if we perform a weak measurement instead. The fidelity between the initial pointer state and the state after the weak measurement interaction $\sigma_w =  \sum_\nu \mel{\chi_\nu}{\rho_i}{\chi_\nu} e^{-i g a_\nu p} \dyad{\phi} e^{ig a_\nu p}$ is given by 

\begin{equation}
F(\sigma_i,\sigma_w) = \sum_\nu \mel{\chi_\nu}{\rho_i}{\chi_\nu} e^{-g^2 a_\nu^2 / (4 \Delta^2)}.
\end{equation}

For the specific case where we measure $\sigma_z$ with the initial system state $\rho_i = \dyad{\uparrow_\xi}$ we found $F(\sigma_i,\sigma_w) = \exp[-g^2/(4\Delta^2)]$. 

Following this weak measurement we perform the post-selection on the system such that the final pointer state is given by $\sigma_{f} = \Tr_S(\Upsilon_f) $, where $\Upsilon_f$ is the total state in Eq. \eqref{eq:total_state_postselection}, and thus the fidelity with the initial state is 

\begin{equation}
\label{eq:fidelity_exact}
F(\sigma_i, \sigma_{f}) = \frac{ \sum_{\nu \mu} \braket{f}{\chi_\nu} \braket{\chi_\mu}{f} \mel{\chi_\nu}{\rho_i}{\chi_\mu} e^{- \frac{g^2}{8 \Delta^2} (a_\nu^2 + a_\mu^2)} }{\sum_{\nu \mu} \braket{f}{\chi_\nu} \braket{\chi_\mu}{f} \mel{\chi_\nu}{\rho_i}{\chi_\mu} e^{- \frac{g^2}{8 \Delta^2} (a_\nu - a_\mu)^2}}.
\end{equation}

This is the exact fidelity between the initial pointer state and the final state after a generalized measurement and post-selection.
The fidelity between the initial state and the final state following the approximate WV derivation in \cite{aharonov1988result} yield $F(\sigma_i,\sigma_{WV} ) = \exp[-g^2(\operatorname{Re} \expval{A}_{WV})^2 / (4 \Delta^2) ]$. 
Note that for the ABL measurement protocol we trivially obtain a vanishing fidelity, $F(\sigma_i, \sigma_{ABL}) = 0$.

Considering the case of measuring $\sigma_z$ with initial state $\rho_i = \dyad{\uparrow_\xi}$ and post-selecting on $\ket{\uparrow_x}$ (corresponding to $\beta = 0$), we found that the exact fidelity \eqref{eq:fidelity_exact} is

\begin{equation}
\label{eq:fidelity_exact_example}
F(\sigma_i, \sigma_f) = \frac{(1 + \cos \alpha)e^{-g^2/(4 \Delta^2)}}{1+ \cos \alpha \, e^{- g^2 / (2\Delta^2)}}.
\end{equation}

\begin{figure}
    \centering
    \includegraphics[width=6.5cm]{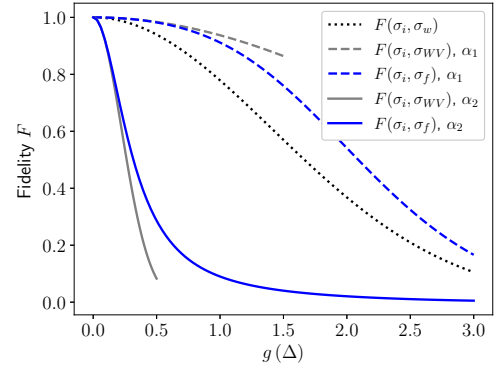}
      \caption{Fidelity of the pointer state for increasing coupling strength $g$. The fidelity of the initial state with the state following a weak measurement and without post-selection $F(\sigma_i,\sigma_w)$ (black, dotted) is independent of the preparation procedure. If a post-selection is performed the fidelity of the initial state with the final pointer state $F(\sigma_i, \sigma_{f})$ is shown for different preparation procedures $\alpha_1 = 0.3 \pi$ (blue, dashed) and $\alpha_2 = 0.9 \pi$ (blue, solid). The fidelity for the WV outcome $F(\sigma_i, \sigma_{WV})$ (gray) only agree with the exact result in the weak coupling regime.}
      \label{fig:fidelity_pointer}
\end{figure}

In Fig. \ref{fig:fidelity_pointer} we show the fidelity \eqref{eq:fidelity_exact_example} for increasing coupling strength $g$ and different preparation procedures $\alpha$. In the limit of $g \ll \Delta$ the fidelity of the WV measurement $F(\sigma_i, \sigma_{WV}) =  \exp[-g^2 \tan^2(\alpha/2)/(4 \Delta^2)]$ is recovered, but we can see that for larger coupling strength the WV is insufficient to describe the change of the pointer state and higher order terms need to be taken into account. In the projective measurement limit $g \gg \Delta $ the fidelity vanish. For comparison we have shown the fidelity without post-selection $F(\sigma_i,\sigma_w) = \exp[-g^2/(4\Delta^2)]$.  

However, we want to emphasize a feature of the pointer fidelity in relation to the performed post-selection. In Fig. \ref{fig:fidelity_pointer} the fidelity between the initial state and the final state after post-selection $F(\sigma_i,\sigma_f)$ is larger than the fidelity without post-selection $F(\sigma_i, \sigma_w)$ for a preparation procedure with $\alpha_1 = 0.3 \pi$, and in contrast $F(\sigma_i, \sigma_f)$ is smaller than $F(\sigma_i, \sigma_w)$ for the preparation angle $\alpha_2 = 0.9 \pi$. The preparation procedure of the initial state $\rho_i(\alpha)$ has a larger overlap with the final post-selected state $\ket{\uparrow_x}$ for $\alpha_1$ or with the orthogonal state $\ket{\downarrow_x}$ for $\alpha_2$.
Thus, for the case of $\bra{\downarrow_x} \rho_i(\alpha) \ket{\downarrow_x} \ge \bra{\uparrow_x}\rho_i(\alpha) \ket{\uparrow_x}$, the initial state $\rho_i(\alpha)$ and final state $\ket{\uparrow_x}$ tend to become orthogonal, leading to a larger WV and hence increased pointer shift. This ultimately reduce the fidelity $F(\sigma_i, \sigma_f)$.
We shall now prove that this property is not accidental for the two angles shown, but holds for a general initial state $\ket{i}$ and final state $\ket{f}$ and corresponding orthogonal state $\ket{f_\perp}$. Essentially, we have to proof that 

\begin{equation}
F(\sigma_i, \sigma_w) \ge F(\sigma_i,\sigma_f),
\end{equation}

if $\abs{\braket{f_\perp}{i}} \ge \abs{\braket{f}{i}}$. Inserting the expressions for the fidelity we find that, $\cos \alpha \, e^{-g^2/(2 \Delta^2)} \ge \cos \alpha$. This is only true if $\cos \alpha \le 0$, implying that $\alpha \in [{\pi}/{2}, {3 \pi}/{2}]$, which complete the proof. Note that this is the same region of initial states where the WV exceeds the eigenvalue range leading to anomalous pointer shifts.

\subsection{Relation to weak value amplification}

We have seen that the outcome of a weak measurement including post-selection can cause the pointer state to shift almost arbitrarily large, leading to an enhancement effect for measuring the small coupling parameter $g$. 
This effect of large pointer shifts due to WV is used in the weak value amplification (WVA) technique, where a small signal can receive huge amplification effects due to post-selection \cite{hosten2008observation}. This amplification is at the expense of a reduced post-selection probability for almost orthogonal initial and final states necessary to obtain large pointer shifts. 
Nevertheless, the WVA technique has been used to increase the measurement sensitivity of small effects in various experiments, and it was shown that WVA has an advantage in terms of quantum metrology by reducing different types of noise and making use of quantum resources \cite{hosten2008observation, dixon2009ultrasensitive, PhysRevLett.125.080501, starling2009optimizing, feizpour2011amplifying,dressel2013strengthening, jordan2014technical}.
Since the demonstrated advantage depends on the assumptions made about the experimental conditions, such as detector saturation or noise, there is still an ongoing debate under which conditions WVA can outperform a conventional measurement scheme \cite{brunner2010measuring, knee2013quantum, tanaka2013information, ferrie2014weak, knee2014amplification,zhang2015precision}. 
However, considering realistic experimental conditions, e.g. including detector saturation and noise, it was conjectured \cite{vaidman2014comment} and shown \cite{harris2017weak}, that WVA can outperform conventional schemes, and can approach the quantum Cramér-Rao bound for parameter estimation \cite{pang2014entanglement, PhysRevLett.125.080501}. 

In the following we want to answer the question which measurement protocol provides the most unambiguous parameter detection, to extract the classical information?
The larger the distinguishability between the initial $\sigma_i$ and final pointer state $\sigma_f$, the higher the probability to unambiguously correlate the measured pointer position to the previous interaction. Thus pointer readings are not mistakenly assigned to the interaction while the pointer is actually obtained from the initial position for large uncertainties.
A measure to quantify this distinguishability between the pointer states is given by $1-F(\sigma_i, \sigma_f)$, implying that it is favorable to minimize the fidelity $F(\sigma_i, \sigma_f)$ in respect of increasing the unambiguous measurement detection.

In general, we observe, that the fidelity of the pointer between the initial and final state $F(\sigma_i,\sigma_f)$ is reduced for larger coupling strength and vanish for a projective measurement ($g \gg \Delta$). For moderate coupling strength $g \simeq \Delta$ the fidelity is reduced when an additional post-selection is performed on an almost orthogonal final state (see Fig. \ref{fig:fidelity_pointer}). 
Thus we assert that a conventional projective measurement is favorable in terms of an unambiguous retrodiction of the pointer shift. 
However, the conditions on the parameters ($g, \Delta$) are usually constraint by the experimental setup. And since the aim of WVA is to measure a very small parameter $g$, the constraint on $\Delta \ll g$ for projective measurements are severe.  
Thus, when taking into account this constraint imposed by the experimental setup the fidelity between initial and final pointer state is non-vanishing. Still pursuing for the least fidelity measure, we find that for an experiment with fixed $g \simeq \Delta$, the largest decrease of the fidelity $F(\sigma_i, \sigma_f)$ is observed when including post-selection on almost orthogonal initial and final states, precisely the requirements of anomalous WV (Fig. \ref{fig:fidelity_pointer}).
We thus observe, that under the possible constraint imposed by the experiment, a weak measurement protocol including post-selection lead to the largest detection sensitivity of the small coupling parameter $g$.

Note that a weak measurement alone does not lead to a pronounced decrease of the fidelity and only in combination with a post-selection on an orthogonal state a significant decrease of the fidelity is observed. 
Thus the unambiguous differentiation between the pointer state before and after the measurement protocol is enhanced by a proper post-selection, in accordance with the observed increase of quantum Fisher information for post-selected weak measurements compared to weak measurements alone \cite{pang2015improving}.

\section{\label{sec:minimal_disturbance}Boundary for minimal measurement disturbance}

We shall now quantitatively analyze the usual assumption that in a WV measurement protocol the disturbance of the initial state is negligible during the intermediate weak measurement \cite{aharonov1990properties}. 
We therefore define a lower bound for the minimal state disturbance of the system in order to extract a desired amount of information in terms of the pointer change.

\subsection{Boundary from pointer perspective}

To quantify the measurement sensitivity of the extracted information from the pointer perspective, we will make use of the pointer fidelity.
The fidelity for the pointer state imply that more information about the observable is extracted for lower values, in the sense that it makes the outcome more distinguishable from the initial state. We thus want to place an upper bound on the pointer fidelity $F_{b}$, such that information about the interaction can unambiguously obtained. Since the fidelity reduces for larger coupling strength, we deduce the minimum coupling strength $g_{min}$, in order to obtain $F_{b}$ for the different measurement protocols. 

For a weak measurement without post-selection with $F_b(\sigma_i, \sigma_w) = \exp[-g^2/(4 \Delta^2)]$, we readily found that $g_{min}^{I} = 2 \Delta \sqrt{\abs{\ln F_{b}}}$. However, if we perform a subsequent post-selection on the system we can solve \eqref{eq:fidelity_exact_example} for $g_{min}$  and obtain  

\begin{equation}
\label{eq:minimal_coupling_postselection}
g_{min}^{II} = 2 \Delta \sqrt{\ln\left[ \frac{C(\alpha) + \sqrt{C(\alpha)^2 - 4 F_{b}^2 \cos \alpha} }{2 F_{b}} \right] },
\end{equation}

where $C(\alpha) = 1+\cos \alpha$. We can now compare the arguments of the square root from $g_{min}^I$, and $g_{min}^{II}$, to identify the regime in which the coupling strength is minimized by the post-selection procedure. 
Assuming that $g_{min}^{II} \le g_{min}^{I}$, imply that, $ \ln F_b \le - \ln [\kappa(\alpha)/(2F_b)]$, where $\kappa (\alpha) = C(\alpha) + \sqrt{C(\alpha)^2 - 4 F_b^2 \cos \alpha}$. This holds only when $\cos \alpha \le 0$, i.e. $\alpha \in [\pi/2, 3 \pi/2]$. Thus, in the regime when initial and final state become orthogonal, then the post-selection procedure can improve the measurement sensitivity. This is to be understood in terms of obtaining the same pointer fidelity $F_b$ as in the case without post-selection for a smaller coupling strength $g_{min}^{II} \le g_{min}^{I}$. 

If we insert these lower bounds for the coupling strength into the trace distance of the system states of Eq. \eqref{eq:distance_initial_stron_vs_weak}, we have a quantitative measure for the actual state disturbance after the intermediate weak measurement. 
We find that for the weak measurement case without post-selection 

\begin{equation}
\label{eq:tradeoff_1}
D_{min}^{I}(\rho_i, \rho_w) = \frac{1}{2} \abs{\cos \alpha} (1- F_{b}^2).
\end{equation}

Using the relation between trace distance and fidelity $D(\sigma_1,\sigma_2) \le \sqrt{1-F(\sigma_1,\sigma_2)^2}$ \cite{fuchs1999cryptographic}, we obtain a lower bound on the state disturbance of the system with respect to the trace distance of the pointer $D_{p}(\sigma_i,\sigma_w)$ 

\begin{equation}
\label{eq:tradeoff_ineq}
D_{min}^{I}(\rho_i, \rho_w) \ge \frac{1}{2} \abs{\cos \alpha } D_p^2(\sigma_i, \sigma_w).
\end{equation}

This is the trade-off relation for the state disturbance of the system state in terms of the pointer state change.

\begin{figure}
    \centering
    \includegraphics[width=6.5cm]{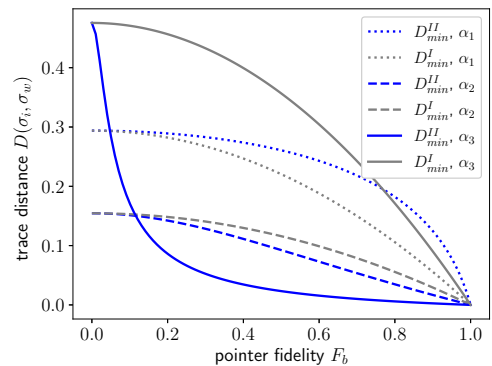}
      \caption{Trace distance of the system states of Eq. \eqref{eq:distance_initial_stron_vs_weak} as a measure of the state disturbance for the measurement interaction strength $g_{min}$ to obtain a desired pointer fidelity $F_b$ for different initial preparations $\alpha_1 = 0.3 \pi$ (dotted), $\alpha_2 = 0.6 \pi$ (dashed) and $\alpha_3 = 0.9 \pi$ (solid). A reduced state disturbance due to post-selection (blue, solid and dashed) is found for $\alpha \in [\pi/2, 3 \pi/2]$ compared to a weak measurement without post-selection (gray). }
      \label{fig:minimal_disturbance}
\end{figure}

However, performing an additional post-selection the minimal coupling strength needed to obtain the same pointer fidelity $F_b$ is different. Using \eqref{eq:distance_initial_stron_vs_weak}  and \eqref{eq:minimal_coupling_postselection} we find for the minimal state disturbance 

\begin{align}
\label{eq:tradeoff_2}
D_{min}^{II}(\rho_i, \rho_w) = \frac{1}{2} \abs{\cos \alpha} \left[ 1- \gamma(\alpha) F_{b}^2 \right],
\end{align}

where $\gamma (\alpha) = 4/\kappa^2(\alpha)$. If $\gamma(\alpha) \ge 1$ the distinguishability between the initial state and the state following the weak measurement is decreased compared to the non-post-selection scenario \eqref{eq:tradeoff_1}. We found that $\gamma(\alpha) = 4 / \kappa^2(\alpha) \ge 1$, if $\cos \alpha \le 0 $, which is the case for $\alpha \in [\pi/2, 3 \pi/2]$, and thus $D_{min}^{II}(\rho_i, \rho_w) \le D_{min}^{I}(\rho_i, \rho_w)$. Hence, a reduced state disturbance due to post-selection is given for the same initial preparation where we obtain anomalous large WV and a reduced pointer fidelity. 
The state disturbance of \eqref{eq:tradeoff_1} and \eqref{eq:tradeoff_2} can be seen in Fig. \ref{fig:minimal_disturbance}, where a considerably large advantage in terms of a reduced disturbance due to post-selection can be found for the case of $\alpha_3 = 0.9 \pi$ (solid). 
Note that technically the post-selection itself not reduces the system state disturbance in the sense that it cures a previous larger disturbance, but it allows to operate the experiment with a coupling strength $g_{min}$ smaller than for the case without post-selection in order to reach the same fidelity for the pointer state $F_{b}$.

This observation is in accordance with the advantage of the weak value amplification technique for small coupling strength. If we wish to measure a small coupling parameter $g$ the measurement sensitivity of the pointer can considerably increase if a post-selection on a system state orthogonal to the initial state is performed.

\subsection{Is the weak measurement disturbance negligible?}

We have seen, that it is necessary to disturb the system in order to extract any information from it \cite{fuchs1996quantum, fuchs1998information, fuchs2001information}, whereas the disturbance of the system in a WV measurement protocol is often neglected \cite{aharonov1990properties}. Thus, it is assumed that $D(\rho_i, \rho_w) \simeq 0$.
Given the lower bound of the coupling strength $g_{min}$ for a desired pointer fidelity, we can ask the question whether the disturbance of the weak measurement is actually negligible or not?

The disturbance of the system for the coupling strength $g_{min}$, needed to obtain the pointer fidelity $F_b$, is given by $D_{min}^{II} (\rho_i, \rho_w)$ (Eq. \eqref{eq:tradeoff_2}) and $D_{min}^{I} (\rho_i, \rho_w)$ (Eq. \eqref{eq:tradeoff_1}), for the case with and without post-selection, respectively. 
For realistic experimental conditions with an initial state of angle $\alpha = 170^\circ$ and post-selecting  for $\beta = 0$, we found that the state disturbance for a pointer fidelity of $F_b = 0.1$ is given by $D_{min}^{II}(\rho_i, \rho_w) = 0.06$. And depending on the desired task, this disturbance can be considered as negligible. 
However, here we have used that following the weak measurement a post-selection on the system is performed such that the minimal coupling strength can be reduced to obtain the same pointer fidelity as opposed to the case of a single weak measurement with-out post-selection in which $g_{min}^{I} \ge g_{min}^{II}$. Consequently the disturbance for the case without post-selection for the same experimental conditions, is $D_{min}^{I}(\rho_i, \rho_w) = 0.49$, and can by no means neglected. 

Thus, for the measurement protocol including post-selection, which can operate with a smaller coupling strength to obtain the same detection sensitivity, we found that the disturbance due to the intermediate weak measurement can be neglected. In contrast, a single weak measurement without post-selection will significantly disturb the system to reach the equivalent measurement sensitivity.

Important for anomalous WV measurements is that the initial $\ket{i} = \cos \alpha/2 \ket{\uparrow_x} + \sin \alpha /2 \ket{\downarrow_x}$, and final $\ket{f} = \ket{\uparrow_x}$ state need to be almost orthogonal, resulting in a reduced signal intensity due to post-selection. Since the state disturbance during the weak measurement is usually neglected, the success probability of the post-selection measurement outcomes is likewise approximated by $P_{ap} = \abs{\bra{f}\ket{i}}^2 = \abs{\cos {\alpha}/{2}}^2$, whereas the exact post-selection probability \eqref{eq:post-selection-probability} is given by, $P_f = \frac{1}{2} (1 + \cos \alpha \, \exp[-g^2/(2 \Delta^2)])$. For the experimental parameters from above with $g_{min}^{II}$ from \eqref{eq:minimal_coupling_postselection}, we found that $P_{ap} = 0.0076$, but for the exact post-selection probability we have $P_f = 0.071$, which exceed the approximated post-selection probability by nearly one order of magnitude.   
The fact that the exact post-selection probability is considerably larger than the approximated overlap of initial and final state, provides further advantage for the weak value amplification technique in the sense that the true number of counts is much larger than often assumed.

\section{\label{sec:conclusion}Conclusions}

We have analyzed the trade-off between the state disturbance of the system and the change of the device state for weak measurements with and without post-selection. In comparison to projective measurements, it holds in general that the distinguishability of the initial system state with the state following a weak measurement is reduced compared to the projective case, i.e. $D(\rho_i, \rho_w) \le D(\rho_i, \rho_s)$, while the same relation holds for the distinguishability of the device states, i.e. $1- F(\sigma_i, \sigma_w) \le 1- F(\sigma_i, \sigma_s)$. Thus the trade-off for an increased information gain of the device is the larger system state disturbance.
Due to the non-perturbative approach we obtained the exact trade-off for the minimal system state disturbance of a weak measurement $D_{min}^{I}$ in terms of the pointer state change. A lower bound for the disturbance in terms of the pointer distinguishability was obtained $D_{min}^{I}(\rho_i, \rho_w) \ge \frac{1}{2} \abs{\cos \alpha} D_p ^2 (\sigma_i, \sigma_w)$. 

However, if a subsequent post-selection is performed, the interaction strength can be considerably decreased while maintaining the same sensitivity in the pointer distinguishability. We found that including post-selection  $D_{min}^{II}(\rho_i,\rho_w) \le D_{min}^{I}(\rho_i, \rho_w)$, if the initial and final state tend to become orthogonal.
The observed advantage of a reduced state disturbance in a weak measurement protocol including post-selection is a collective effect of the preserved coherence and corresponding minimal disturbance during the intermediate weak measurement with the fact that when performing a proper post-selection a smaller coupling strength is sufficient to obtain the same pointer change.
The fact that this combined effect of weak measurement and post-selection lead to a larger pointer shift in the case of nearly orthogonal post-selection can significantly outperform a weak measurement alone. This observation is therefore directly connected to the appearance of anomalous weak values.

The analysis of quantum information measures and measurement disturbance trade-off relations in weak measurement schemes are of current interest \cite{cheong2012balance, fan2015trade} and are particularly related to quantum metrology analysis, where is was recently shown that the Fisher information of the measurement outcome can be increased in the post-selected sub-ensemble \cite{PhysRevLett.125.080501, arvidsson2019quantum}. 
It would therefore be of direct interest if the shown advantage of imaginary weak values for quantum metrology \cite{kedem2012using} is likewise observed in the measurement disturbance trade-off relation. 
We further believe that an extended analysis for open quantum systems \cite{ivanov2020feedback, schlosshauer2020protective} or the use for quantum communication protocols \cite{anwer2020experimental} are of current interest. \\


\bibliographystyle{unsrt}
\bibliography{literatur}{}

\begin{thebibliography}{10}
\bibitem{note2}
Fixing the angle $\beta = 0$ does not change the behaviour of the measurement
  outcomes as long as $\beta \neq \pi/2, \, 3 \pi/2 $, where
  $\ket{\uparrow_\eta}$ is an eigenstate of $\sigma_z$.
\end{thebibliography}
\end{document}